\documentclass[a4paper,11pt]{amsart}
\usepackage{amsmath,amssymb,amsfonts,phonetic}
\usepackage{epsfig}
\usepackage{graphicx}
\usepackage{mathrsfs}

\newtheorem{thm}{Theorem}[section]

\newtheorem{rmk}{Remark}[thm]
\newtheorem{lma}{Lemma}[section]

\newtheorem{corl}{Corollary}[section]
\def\N{{\rm I\kern-0.16em N}}
\def\R{{\rm I\kern-0.16em R}}
\def\E{{\rm I\kern-0.16em E}}
\def\P{{\rm I\kern-0.16em P}}
\def\F{{\rm I\kern-0.16em F}}
\def\B{{\rm I\kern-0.16em B}}
\def\C{{\rm I\kern-0.46em C}}
\def\G{{\rm I\kern-0.50em G}}
\numberwithin{equation}{section}
\font\eka=cmex10
\usepackage{ae}
\def\ind{\mathrel{\hbox{\rlap{%
\hbox to 7.5pt{\hrulefill}}\raise6.6pt\hbox{\eka\char'167}}}}
\begin{document}

\title [ GFBM-- market with transaction costs]
{ On the fractional Black-Scholes market with transaction costs}

\author[Azmoodeh]{Ehsan Azmoodeh}

\address{ Department  of Mathematics and Systems Analysis\\ Aalto University\\
P.O. Box 11100, 00076 Aalto, Finland}

\email{azmoodeh@cc.hut.fi}

\begin{abstract}
We consider fractional Black-Scholes market with proportional transaction costs. When transaction costs are present, one trades periodically i.e. we have the discrete trading with equidistance $n^{-1}$ between trading times. We derive a non trivial hedging error for a class of European options with convex payoff in the case when the transaction costs coefficients decrease as $n^{-(1-H)}$. We study the expected hedging error and asymptotic behavior of the hedge as $H \to \frac{1}{2}$.
\medskip


\smallskip


\smallskip


\end{abstract}

\maketitle
\section{Introduction}
Absence of arbitrage is a weak form of financial equilibrium and it plays a fundamental role in mathematical finance. In fractional Black-Scholes market, stock price is modelled by geometric fractional Brownian motion. This model admits arbitrage opportunities without transaction costs with continuous trading. Guasoni \cite{g} showed that, if one considers such financial markets with proportional transaction costs, then arbitrage opportunities disappear. In \cite{g-r-s}, the authors studied a more general formulation of such financial markets with proportional transaction costs and they also find a superreplication price for a class of European options.\\

Leland \cite{le} suggests a way to include proportional transaction costs in classical Black-Scholes market. Namely, one trades in discrete and equidistant trading times i.e. periodically. Over each trading subinterval, the trader follows the delta hedging strategy computed at the left point of the trading subinterval with a modified volatility. The modified volatility depends on the original volatility and the number of trading intervals (see \cite{k-s}). Leland remarked that the price of the modified strategy approximately hedges the option payoff at terminal date as the length of trading intervals tends to zero.\\

The Leland's approach is considered useful for practitioners, although the modified strategy does not provide an exact hedge in the case when the level of transaction costs is constant. First Lott \cite{lo} and later Kabanov and Safarian \cite{k-s} showed that this is true when transaction cost coefficients $k_n$ decrease as 
\begin{equation*}
 k_n = k_0 n^{- \alpha}, \qquad \alpha \in ( 0 , \frac{1}{2} ],
\end{equation*}
where $\frac{1}{n}$ is the length of the trading intervals. Constant transaction costs correspond to the case $\alpha = 0 $. Kabanov and Safarian \cite{k-s} computed the hedging error explicitly for European call option and in general it is not equal to zero. They showed that the limiting hedging error is positive with probability one. This means that the    option is always underpriced in the limit.\\

The motivation of this study comes from the recent work by Azmoodeh et. al. \cite{a-m-v}. There we studied a hedging problem for European options with convex payoff in fractional Black-Scholes market with Hurst parameter $H > \frac{1}{2}$. We assume that the market is frictionless and continuous trading is possible. Then any European option with convex payoff can be hedged perfectly. Moreover, hedging strategy and \textit{hedging cost} (see \cite{b-s-va}) are given explicitly. Simply speaking we showed that the classical chain rule holds for convex functionals of geometric fractional Brownian motion. Moreover, the wealth process, which is stochastic integral of replication portfolio with respect to stock price process $S$, is the limit of Riemann-Stieltjes sums almost surely. This makes our model more interesting from financial point of view (see \cite{b-h} and \cite{s-v}).\\

European call option with strike price $K$  serves as a motivating example for us. In this case the \textit{stop-loss-start-gain} strategy 
\begin{equation*}
u_t = 1_{ \{ S_t > K \}}
\end{equation*}
is self-financing replication strategy in our model. Note that this replicating strategy is not self-financing in the classical Black-Scholes market with standard Brownian motion. This strategy is of unbounded variation, and hence it is not practical for our model with proportional transaction costs and continuous trading. One possibility is to trade periodically, and the level of transaction costs is the function 
\begin{equation*} 
 k= k_n = k_0 n ^{-(1-H)}
\end{equation*}
of the number of trading intervals. This is similar to Leland \cite{le} in the case of classical Black-Scholes model with transaction costs.\\

The paper is organized as follows. Section $2$ includes some auxiliary results to handle the transaction costs term in the limit. Also the description of the model in precise way is given. Section $3$ contains the main result. The paper ends with more discussion on specific example European call option and conclusion in the section $4$.\\

\section{Preliminaries}

Throughout the paper $(\Omega,\mathcal{F},\mathbb{P})$ stands for a complete probability space. Assume $ B^{H} =( B^{H}_t ) _{t \in [0,T]}$ be a standard fractional Brownian motion with Hurst parameter $ H \in (0,1)$ and $ S_t = S_0 \exp\{B^{H}_t \}$ be geometric fractional Brownian motion, $S_0 \in \mathbb{R_{+}}$. 

\subsection{Auxiliary results on convex functions}

We recall some results on convex functions. First, recall that
every convex function $f:\mathbb{R} \to \mathbb{R}$ has a left-derivative $f'_-$ and a right-derivative $f'_+$.\\

The next theorem gives information about the left-derivative $f'_-$  and right-derivative $f'_+$.

\begin{thm}\cite{r-y} \label{thm:Yor1}
The functions $f^{'}_{-}$ and $ f^{'}_{+}$ are increasing, respectively
left and right-continuous and the set $\{ x: f^{'}_{-}(x) \neq f^{'}_{+}(x) \}$ is at most countable.
\end{thm}

Moreover, the second derivative of a convex function $f$ exists as a distribution, and first derivative can be represented in terms of the second derivative.

\begin{thm}\cite{r-y}
 The second derivative $f^{''}$ of convex function $f$ exists in the sense of distributions,
and it is a positive Radon measure; conversely, for any Radon measure
$\mu$ on $\mathbb{R}$, there is a convex function $f$ such that $f^{''} = \mu$
and for any interval $I$ and $x \in \mbox{\rm int} (I)$ we have the equality
\begin{equation}\label{eq:first}
f^{'}_{-} (x) = \frac{1}{2} \int_{I} \mbox{\rm sgn} (x-a) \mu (da)+ \alpha _I,
\end{equation}
where $\alpha_I$ is a constant and $\mbox{\rm sgn } x = 1 $ if $x>0$ and $-1$ if $x\le 0$.
 \end{thm}

\begin{rmk}\label{rmk:Yor2}
 If the $\mbox{supp}(\mu) $ is compact, then one can globally state that
\begin{equation}\label{eq:second}
f^{'}_{-} (x) = \frac{1}{2} \int \mbox{\rm sgn} (x-a) \mu (da)
\end{equation}
up to a constant term.
\end{rmk}
\vskip0.10cm

\subsection{Linear approximation of convex functions}

Let $f:\mathbb{R} \to \mathbb{R} $ be a convex function. For each interval $[a,b]$, let $\pi = \{ a=a_0 < a_1 < ... < a_n=b \} $ be a partition of the interval and 
\begin{equation*}
\Vert \pi \Vert:= \max _{1 \le i \le n} (a_i - a_{i-1}).
\end{equation*}
A piecewise linear function through points $(a_i , f(a_i))$  is called a \textit{convex linear approximation} of convex function $f$ on the interval $[a,b]$ based on the partition $\pi$.

\begin{thm}\label{thm:approx}
Let $[a,b]$ be a closed interval and $\{ \pi _m \}$ be a sequence of partitions of the interval $[a,b]$ such that
\begin{equation*}
\Vert \pi_m \Vert \to 0  \quad \text{as} \quad m \to \infty 
\end{equation*}
where $\pi_m= \{ a_1,a_2,...,a_{n(m)} \}$.

Let $P_{m}$  be a convex linear approximation of convex function $f$ on the interval $[a,b] $ based on partitions $ \pi _m $. Then we have
\begin{itemize}
\item[(i) ]
On the interval $(a,b)$ as $ m \to \infty$
\begin{equation*}
 P_m \to f \qquad \text{and} \qquad (P_m)^{'}_{-} \to f^{'}_{-}  \qquad \text{pointwise} .\\
\end{equation*}  
\item[(ii) ]
For any bounded continuous function $g$ we have 
\begin{equation*}
\int_{[a,b]} g d\mu _{m} \to \int_{[a,b]} g d \mu \qquad \text{as}\quad m  \to \infty ,
\end{equation*}
where $\mu _m$ stands for Radon measure corresponding to the second derivative of $P_m$.\\
\end{itemize}
\end{thm}

\subsection{Approximation of local time of fractional Brownian motion}
\vskip0.10cm
The \textit{occupation measure} related to fractional Brownian motion is defined by
\begin{equation*}
 \Gamma _{B^H} (I \times U) = \lambda \{ t \in I : B^{H}_t \in U \} = \int_{ I} \textbf{1}_{ \{ B^H _{t} \in U \} } dt
\end{equation*}
where $I$ and $U$ are Borel sets on time interval $[0,T]$ and the real line respectively and $\lambda$ stands for Lebesgue measure. It is well-known that the occupation measure has a jointly continuous density (\textit{local time}) which is denoted by $l^{H} (x,t) := l^{H} (x,[0,t])$ and is H\"older continuous in $t$ of any order $\alpha < 1-H$  and in $x$ of any order $\beta < \frac{1-H}{2H}$ (for a survey article on the subject see Geman and Horowitz \cite{g-h}).\\

Let $ B^{H}_{\Delta} = \{ B^{H}_{\Delta} (t) \}_{t \in [0,T]}$ be polygonal approximation of size $\Delta$ of $B^{H}$ i.e. $ B^{H}_{\Delta}$ is the polygonal lines which connect points $ \{ (i \Delta, B^{H}(i \Delta) ) \}$ for suitable running index $i$. Set
\begin{equation*}
 C^{a}_{\Delta} (B^H, [0,T])= \{ t \in [0,T]: B^H _{\Delta} (t) = a \quad \text{and} \quad t \neq i\Delta \quad \text{for each index }\, i \}
\end{equation*}
and
\begin{equation*}
 N^{a}_{\Delta}(B^H, [0,T])= \# \, C^{a}_{\Delta} (B^H, [0,T]), 
\end{equation*}
i.e. the number of level $a$ crossing of $B^H _{\Delta} $ over interval $[0,T]$. Then we have the following approximation for the local time $l^{H} (a,t)$.

\begin{thm} \label{thm:azais}
 Assume $ B^{H}$ be a standard fractional Brownian motion with $ H \in (0,1)$ and $ N^{a}_{\Delta}(B^H, [0,T])$ be the number of level $a$ crossing of size $ \Delta -$ polygonal approximation of $B^H$. Then 
\vskip0.25cm
\begin{equation*}
 \sqrt{\frac{\pi}{2}} \Delta ^ {1-H}  N^{a}_{\Delta}(B^H, [0,T]) \to l^H  (a,[0,T])\quad \text{in} \quad  L^2 \quad \text{ as } \; \Delta \to 0.
\end{equation*}
\end{thm}
\textbf{Proof}. See \cite{a}, Theorem $5$.

\subsection{The model}
Let $f: \mathbb{R} \to \mathbb{R}$ be a convex function with positive Radon measure $\mu$ corresponding to its second derivative. We showed in \cite{a-m-v} that in a geometric fractional Brownian motion market model with $H>\frac{1}{2}$, all European options with the terminal payoff $ f(S_T) $ can be hedged perfectly and the hedging strategy and hedging cost are  $f^{'}_{-} (S_t)$ and $f(S_0)$ respectively. More precisely we have that 

\begin{equation}\label{eq:chain}
 f(S_T) = f(S_0) + \int _0 ^T f^{'}_{-} (S_t) d S_t.
\end{equation}
 The stochastic integral in the right-hand side is understood as a limit of Riemann--Stieltjes sums almost surely, i.e.

\begin{equation}\label{eq:RSsum}
\sum_{i=0}^{n} f^{'}_{-}  (S_{t^{n}_{i-1}}) (S_{t^{n}_{i}} - S_{t^{n}_{i-1}})     \stackrel{\text{a.s}}{\longrightarrow} \int_{0}^{T} f^{'}_{-}  (S_{t}) dS_t, \qquad t^{n}_{i}=\frac{iT}{n}.
\end{equation}

Assume that the terminal trading time $T=1$. For each $n$ we divide the trading interval $[0,1]$ to $n$ subintervals $[t^{n}_{i-1}, t^{n}_i]$ where 
\begin{equation*}
 t^{n}_i =\frac in = i \Delta_n ,\qquad i=0,1,...,n.
\end{equation*}
The two-assets market model consists of :\\
\begin{itemize}
\item[(i)] Riskless asset (bond), $ B_t = 1; \quad t \in [0,1]$ which corresponds to zero interest rate.\\
\item[(ii)] Risky asset (stock) whose price is modeled by geometric fractional Brownian motion 
\begin{equation*}
 S_t = S_0e^{B^H _t }; \quad t \in [0,1], \quad \text{and} \quad H>\frac{1}{2}.
\end{equation*}
\end{itemize}

Now consider the \textit{discretized version} of hedging strategy, i.e. 
\begin{equation*}
\theta _t ^n = \sum_{i=1}^{n} f^{'}_{-}(S_{t^{n}_{i - 1}}) 1_{(t^{n}_{i-1},t^{n}_i]}(t);\qquad t\in(0,1].
\end{equation*}

In the presence of proportional transaction costs, the value of this portfolio at terminal date is

\begin{equation}\label{eq:value-process}
V_1 (\theta ^n) = f(S_0) + \int _{0}^{1} \theta _t ^n dS_t - k \sum_{i=1}^{n}S_{t^{n}_{i-1}} |f^{'}_{-}( S_{t^{n}_{i}}) - f^{'}_{-}( S_{t^{n}_{i - 1}} )|.
\end{equation}

Here it is assumed that the transaction costs are "two-sided" i.e. buying and selling are equally charged and transaction costs coefficient $k$ is a function of the length of trading intervals, i.e.
\vskip0.10cm
\begin{equation*}
k= k_n = k_0 n ^{-(1-H)}, \quad k_0 > 0.
\end{equation*}
\begin{rmk}
In general the hedging strategy $f^{'}_{-} (S_t)$ is not of bounded variation. The strategy 
\begin{equation*}
 f^{'}_{-} (S_t)= 1 _{\{S_t > K \}}
\end{equation*}
is of unbounded variation, for $f(x)=(x-K)^+$ corresponding to European call option with strike price $K$.
\end{rmk}

\begin{rmk}
Note that it is assumed that there is no transaction costs at time $t=0$ when the trader enters to the market.
\end{rmk}

\section{Main Result}
After the preliminaries and the precise description of the market model, we can state the main result of the paper.

\begin{thm}
Assume the level of transaction costs $k= k_n = k_0 n ^{-(1 - H)}$, where $k_0 > 0$. Then
\begin{equation*}
\P \hbox{-} \lim_{n \to \infty} V_1( {\theta}^{n} ) = f(S_1) - \mathbf{J},
\end{equation*}
where 
\begin{equation*}
  \mathbf{J}= \mathbf{J}(k_0):= \sqrt{\frac{2}{\pi}} k_0 \int_{\R } \int_0 ^ 1 S_t l^{H}(\ln a ,dt) \mu (da),
\end{equation*}
and the inner integral in the right hand side is understood as limit of Riemann-Stieltjes sums a.s.
\end{thm}

When the number of portfolio revision increases fast enough or equally saying, the transaction costs coefficients $k_n$ decrease faster, we have the perfect replication in the limit, i.e. we have the following result.

\begin{corl} Let the level of transaction costs $k= k_n = k_0 n ^{-\alpha}$ where $\alpha > 1 - H $, $k_0 > 0$. Then
 \begin{equation*}
\P \hbox{-} \lim_{n \to \infty} V_1( {\theta}^{n} ) = f(S_1).
\end{equation*}
\end{corl}

\textbf{Proof}: We can assume that the support of $\mu$ is compact otherwise one can consider auxiliary convex functions

\begin{equation*}
 \tilde{f}_n (x)=
\begin{cases}
 f^{'}_{+}(0)x + f(0),& \text{if $ x<0$},\\
f(x),& \text{if $0 \le x \le n $},\\
f^{'}_{-}(n)(x-n) + f(n),& \text{if $ x > n$}.
\end{cases}
\end{equation*}
(see \cite{a-m-v} for details). Since

\begin{equation*}
\left \{ \begin{aligned}
f(S_1) &= f(S_0) + \int _0 ^1 f^{'}_{-}( S_t) d S_t, \\ 
V_1( {\theta}^{n} ) &= f(S_0) + \sum_{i=1}^{n}  f^{'}_{-}( S_{t^{n}_{i - 1}}) (S_{t^{n}_{i}} - S_{t^{n}_{i-1}})  - k_n \sum_{i=1}^{n}S_{t^{n}_{i-1}} |f^{'}_{-}( S_{t^{n}_{i}} ) - f^{'}_{-}( S_{t^{n}_{i - 1}} )|.
\end{aligned}
\right.
\end{equation*}

Therefore, we have the indentity

\begin{equation*}
 V_1( {\theta}^{n} ) - f(S_1) = I^1 _n - k_0  I^2 _n,
\end{equation*}

where 

\begin{equation*}
\left \{ \begin{aligned}
I^1 _n &=  \sum_{i=1}^{n} f^{'}_{-}( S_{t^{n}_{i - 1}})(S_{t^{n}_{i}} - S_{t^{n}_{i-1}})  - \int _0 ^1 f^{'}_{-}( S_t ) d S_t, \\
I^2 _n &= \Delta_n ^{1-H} \sum_{i=1}^{n}S_{t^{n}_{i-1}} |f^{'}_{-}( S_{t^{n}_{i}}) - f^{'}_{-}( S_{t^{n}_{i - 1}})|.
\end{aligned}
\right.
\end{equation*}
\vskip0.25cm

Note that $I^{1}_n \to 0 $ almost surely by (\ref{eq:RSsum}). It remains to study the behavior of the second term $I^2 _n$. The representation (\ref{eq:second}) relates the left derivative of convex function in term $I^2 _n$ to Radon measure of its second derivative. We divide the proof in three steps, depending on the $\mbox{\rm supp } \mu$.
\vskip0.25cm

\textbf{Step 1:} $\mbox{\rm supp } \mu = \{ a \}$.\\ 
We can assume that $\mu (a)= 1 $ and $f^{'}_{-}(x) = 1_{ \{ x>a \} } $. This follows from representation (\ref{eq:second}).

For any $m \ge n$,
\begin{multline*}
\Big|  \Delta_{m}^{1-H} \sum_{j=1}^{m} S_{t^{m}_{j-1}}  |1 _{\{ S_{t^{m}_{j}} > a\}}-1_{\{ S_{t^{m} _{j-1}} > a\}}| - \int_0 ^ 1 S_t l^{H}(\ln a,dt) \Big| \le \\
 \Delta_{m}^{1-H} \Big| \sum_{j=1}^{m} S_{t^{m}_{j-1}}  |1 _{\{ S_{t^{m}_{j}} > a\}}-1_{\{ S_{t^{m} _{j-1}} > a\}}| - \sum_{i=1}^{n} S_{t^{n}_{i-1}} \sum_{j \in I(i)}  |1 _{\{ S_{t^{m}_{j}} > a\}}-1_{\{ S_{t^{m} _{j-1}} > a\}}| \Big| \\
 + \Big| \Delta_{m}^{1-H} \sum_{i=1}^{n} S_{t^{n}_{i-1}} \sum_{j \in I(i)}  |1 _{\{ S_{t^{m}_{j}} > a\}}-1_{\{ S_{t^{m} _{j-1}} > a\}}| -  \sum_{i=1}^{n} S_{t^{n}_{i-1}} l^H (\ln a,(t^{n}_{i-1},t^{n}_{i}]) \Big| \\
+ \Big| \sum_{i=1}^{n} S_{t^{n}_{i-1}} l^H (\ln a,(t^{n}_{i-1},t^{n}_{i}])  - \int_{0}^{1} S_t l^{H}(\ln a ,dt) \Big| \\
= A_{n,m} + B_{n,m} + C_n,\\
\end{multline*}
where for each $i= 1, 2 ,.., n, I(i)= \{ j : t^{m}_{j} \in (t^{n}_{i-1},t^{n}_{i}] \}. $\\
Obviously, $\lim _{n \to \infty} C_n = 0 $, since $l^H (\ln a,\cdot) $ is increasing in $t$ and
\begin{equation*}
l^H (x,(s,t]) = l^H (x,t) - l^H (x,s), \quad s<t.
\end{equation*}
\vskip0.10cm

\begin{multline*}
 |B_{n,m} | \le \sum_{i=1}^{n} S_{t^{n}_{i-1}} \Big|  \Delta_{m}^{1-H} \sum_{j \in I(i)}  |1 _{\{ S_{t^{m}_{j}} > a \}}-1_{\{ S_{t^{m} _{j-1}} > a \}}|- l^H (\ln a (t^{n}_{i-1},t^{n}_{i}]) \Big|.
\end{multline*}
Therefore for each fixed $n$ by Theorem $\ref{thm:azais}$ as $m \to \infty$,
\begin{equation*}
\begin{split}
 \P \hbox{-} &\lim_{m \to \infty} \Big|  \Delta_{m}^{1-H} \sum_{j \in I(i)}  |1 _{\{ S_{t^{m}_{j}} > a \}}-1_{\{ S_{t^{m} _{j-1}} > a \}}|- l^H (\ln a,(t^{n}_{i-1},t^{n}_{i}]) \Big| \\
&= \P \hbox{-} \lim_{m \to \infty} \Big| \Delta_{m}^{1-H} N^{\ln a}_{\Delta _m}(B^H,(t^{n}_{j-1}, t^{n}_{j}]) - l^H (\ln a ,(t^{n}_{i-1},t^{n}_{i}]) \Big|= 0.\\
\end{split}
\end{equation*}
Hence $B_{n,m}$ converges in probability to zero as $m$ tends to infinity.\\

\vskip0.25cm
\begin{equation*}
\begin{split}
 \vert A_{n,m} &\vert \le \sum_{i=1}^{n} \Delta_{m}^{1-H} \sum_{j \in I(i)} \vert S_{t^{n}_{i-1}} - S_{t^{m}_{j-1}} \vert  |1 _{\{ S_{t^{m}_{j}} > a \}}-1_{\{ S_{t^{m} _{j-1}} > a \}}|  \\
&\le \sum_{i=1}^{n} \sup_{u \in (t^{n}_{i-1},t^{n}_{i})} \vert S_{t^{n}_{i-1}} - S_u \vert \Delta_{m}^{1-H}\sum_{j \in I(i)}|1 _{\{ S_{t^{m}_{j}} > a \}}-1_{\{ S_{t^{m} _{j-1}} > a \}}|  \\
&\stackrel{\P}{\longrightarrow} \sum_{i=1}^{n} \sup_{u \in (t^{n}_{i-1},t^{n}_{i})} \vert S_{t^{n}_{i-1}} - S_u \vert l^H (\ln a ,(t^{n}_{i-1},t^{n}_{i}]) =A_n.\\
\end{split}
\end{equation*}

Fix $\varepsilon > 0$. Since $S_t$ is uniformly continuous on compact intervals $[t^{n}_{i-1},t^{n}_{i}] $, there exists  $n_0 \in \mathbb{N}$ such that for all $n \ge n_0$ we have 
\begin{equation*}
 \sup_{u \in (t^{n}_{i-1},t^{n}_{i})} \vert S_{t^{n}_{i-1}} - S_u \vert < \varepsilon, \qquad 1\le i \le n.
\end{equation*}
Therefore, for all $n \ge n_0$,
\begin{equation*}
 A_n \le  \varepsilon  \sum_{i=1}^{n} l^H (\ln a ,(t^{n}_{i-1},t^{n}_{i}]) = \varepsilon  l^H (\ln a , [0,1]).
\end{equation*}
So $A_n $ converges to zero almost surely as $n$ tends to infinity.\\
\vskip0.25cm

\textbf{Step 2:} $\mbox{\rm supp }\mu = \{ a_1, a_2, ..., a_l \}$.\\

Before to show the convergence in this case we need the following simple lemma.
\begin{lma}\label{lma:sum}
 For $x, y \in \mathbb{R}$ and positive numbers $\alpha _1, \alpha _2, ..., \alpha_l$ we have
\begin{equation*}
 |\sum_{j=1}^{l} ( 1_{ \{ y > a_j \} } - 1_{ \{ x > a_j \} }) \alpha_j| = \sum_{j=1}^{l} | 1_{ \{ y > a_j \} } - 1_{ \{ x > a_j \} }| \alpha_j.
\end{equation*}
\end{lma}

\textbf{Proof}: First note that $ 1_{ \{ y > a_j \} } - 1_{ \{ x > a_j \} } = \pm 1,0 $. Assume $y > x $, so this implies that
all coefficients $ 1_{ \{ y > a_j \} } - 1_{ \{ x > a_j \} }$ for $\alpha_j$'s are equal to $1$ or $0$ for $1 \le j \le l$. Hence
\begin{equation*}
\begin{split}
|\sum_{j=1}^{l} ( 1_{ \{ y > a_j \} } - 1_{ \{ x > a_j \} }) \alpha_j|&= \sum_{j=1}^{l} ( 1_{ \{ y > a_j \} } - 1_{ \{ x > a_j \} }) \alpha_j \\
& = \sum_{j=1}^{l} |( 1_{ \{ y > a_j \} } - 1_{ \{ x > a_j \} })| \alpha_j.\\
\end{split}
\end{equation*}
 The case $ x > y$ is similar.\\

Next we work with $ I^2 _n$. 
\begin{equation*}
\begin{split}
I^2 _n &= \Delta ^{1-H}_{n}  \sum_{i=1}^{n}S_{t^{n}_{i-1}} |f^{'}_{-}( S_{t^{n}_{i}} ) - f^{'}_{-}( S_{t^{n}_{i - 1}} )|\\
&= \Delta ^{1-H}_{n} \sum_{i=1}^{n}S_{t^{n}_{i-1}} \Big| \frac{1}{2} \sum_{j=1}^{l} \big[ ( 1_{ \{ S_{t^{n}_{i}}  > a_j \} } - 1_{ \{ S_{t^{n}_{i}}  <  a_j \} }) - (1_{ \{ S_{t^{n}_{i-1}}  > a_j \} } - 1_{ \{ S_{t^{n}_{i-1}}  <  a_j \} } ) \big] \mu (a_j) \Big|\\
&= \Delta ^{1-H}_{n} \sum_{i=1}^{n}S_{t^{n}_{i-1}} \Big| \frac{1}{2} \sum_{j=1}^{l} ( 1_{ \{ S_{t^{n}_{i}}  > a_j \} } - 1_{ \{ S_{t^{n}_{i-1}}  >  a_j \} }) \mu (a_j) \Big| \\
&+ \Delta ^{1-H}_{n} \sum_{i=1}^{n}S_{t^{n}_{i-1}} \Big| \frac{1}{2} \sum_{j=1}^{l} ( 1_{ \{ S_{t^{n}_{i-1}}  <  a_j \} } - 1_{ \{ S_{t^{n}_{i}}  < a_j \} }) \mu (a_j)\Big|\\
&= A_n + B_n \\
\end{split}
\end{equation*}
By lemma \ref{lma:sum} we see that
\begin{equation*}
\begin{split}
A_n &= \frac{1}{2} \Delta ^{1-H}_{n} \sum_{i=1}^{n}S_{t^{n}_{i-1}} \sum_{j=1}^{l} \big|     1_{ \{ S_{t^{n}_{i}}  > a_j \} } - 1_{ \{ S_{t^{n}_{i-1}}  >  a_j \} }     \big|\mu(a_j)\\
&=\frac{1}{2} \Delta ^{1-H}_{n}\sum_{j=1}^{l}\mu(a_j)  \sum_{i=1}^{n}S_{t^{n}_{i-1}}\big|     1_{ \{ S_{t^{n}_{i}}  > a_j \} } - 1_{ \{ S_{t^{n}_{i-1}}  >  a_j \} }     \big|\\
&\stackrel{\P}{\longrightarrow} \frac{1}{2}\sum_{j=1}^{l}\mu(a_j) \int_0 ^ 1 S_t l^{H}(\ln a_j ,dt) = \frac{1}{2}  \int_{\R } \int_0 ^ 1 S_t l^{H}(\ln a ,dt) \mu (da).\\
\end{split}
\end{equation*}
By a similar argument for the term $B_n$ we conclude that 
\begin{equation*}
 \P \hbox{-} \lim_{n \to \infty}I^2 _n = \int_{\R } \int_0 ^ 1 S_t l^{H}(\ln a ,dt) \mu (da).
\end{equation*}
\vskip0.25cm

\textbf{Step 3:} general case. \\

Let compact interval $[a,b]$ contain the $\mbox{ \rm supp }\mu$ and let $P_m$ be the convex linear approximation of convex function $f$ on the interval $[a,b]$ based on equidistant partition of interval $[a,b]$ i.e. polygonal with vertices $\{ \big( a+i\Delta_m (b-a), f(a+i \Delta_m(b-a)) \big) \} _{i=0}^{m}$. Define $P_m$'s the same as $f$ outside the interval $[a,b]$. Note that outside of the interval $[a,b]$ the convex function $f $ is linear. Then for $m,n \in \N $ we have 
\vskip0.10cm
\begin{equation*}
 |I^2 _n - \mathbf{J}|  \le |I^2 _n - I_{n,m}|+ | I_{n,m} - I_m | + | I_m - \mathbf{J}|,
\end{equation*}
where

\begin{equation*}
\left \{ \begin{aligned}
I_{n,m} &= \Delta_n ^{1-H} \sum_{i=1}^{n} S_{t^{n}_{i-1}} \big| (P_{m})^{'}_{-} (S_{t^{n}_{i}} ) - (P_m)^{'}_{-} (S_{t^{n}_{i-1}} ) \big|, \\
I_m &= \int_{\R} \int_{0}^{1} S_t l^{H}(\ln a ,dt) \mu_m (da).
\end{aligned}
\right.
\end{equation*}
By the elementary inequality $||a| - |b| | \le |a-b|$, we have that
\begin{multline*}
|I^2 _n - I_{n,m}| \le \Delta_n ^{1-H} \sum_{i=1}^{n} \Big| \big( P_{m})^{'}_{-} (S_{t^{n}_{i}}) - f^{'}_{-}( S_{t^{n}_{i}}) \big) - \big( P_{m})^{'}_{-} (S_{t^{n}_{i-1}}) - f^{'}_{-}( S_{t^{n}_{i-1}}) \big) \Big|.
\end{multline*}
Now for fixed $n$, by Theorem \ref{thm:approx} the right-hand side converges to zero almost surely as $m$ tends to infinity. By simple calculations
\begin{equation*}
\begin{split}
\int_{0}^{1} S_t l^{H}(\ln a,dt) &=  \int_{0}^{1} S_t \textbf{1}_{ \{ B^{H}_t = \ln a \}} dt \\
& =  \int_{0}^{1} e^{\ln a} \textbf{1}_{ \{ B^{H}_t = \ln a \}} dt \\
&= a l^{H}(\ln a,[0,1]).
\end{split}
\end{equation*}
It follows that 
\begin{multline*}
|I_m - \mathbf{J} | \le \big|\int_{\R} a l^H (\ln a  , [0,1]) \mu_m (da) - \int_{\R} a l^H (\ln a  , [0,1]) \mu(da) \big|.
\end{multline*}
By Theorem \ref{thm:approx} the right-hand side converges to zero almost surely as $m$ tends to infinity, since support of $\mu$ is compact.\\
\vskip0.10cm

Finally, for fixed $m$ by Step 2, we have that 
\begin{equation*}
 \P \hbox{-} \lim_{n \to \infty} I_{n,m} = I_m. 
\end{equation*}

\begin{rmk}
 Clearly, the hedging error
\begin{equation*}
\begin{split} 
\mathbf{J}=\mathbf{J} (k_0): & = \sqrt{\frac{2}{\pi}}k_0 \int_{\R} \int_{0}^{1} S_t l^{H}(\ln a,dt)  \mu(da)\\
&= \sqrt{\frac{2}{\pi}}k_0 \int_{\R} a l^{H}(\ln a,[0,1]) \mu(da)\\
\end{split}
\end{equation*}
is positive a.s. and strictly positive on a set of positive probability. So with proportional transaction costs, the discretized replication strategy  asymptotically subordinates rather than replicate the value of convex European option $f(S_1)$ and the option is always subhedged in the limit. 
\end{rmk}

\begin{rmk}
 The limiting hedging error $\mathbf{J}=\mathbf{J} (k_0)$ is small for small values of fixed proportional transaction costs coefficient $k_0$.
\end{rmk}

\section{Discussion and conclusion}
\subsection{The case European call option}
Consider European call option with corresponding convex function $f(x)=(x-K)^+$. The approximating wealth process can not hedge perfectly the option payoff $(S_1 - K )^+$ and limitting hedging error takes the form 
\begin{equation*}
\mathbf{J} = K l^{H}(\ln K , [0,1]).
\end{equation*}
So, it is interesting to examine the expected hedging error 
\begin{equation*}
 \begin{split}
  \E (\mathbf{J}) & = K \E \big( l^{H}(\ln K , [0,1]) \big) \\
& = \frac{K}{\sqrt{2 \pi}} \int_{0}^{1} t^{-H} \exp \{- \frac{1}{2} t^{-2H} \ln ^{2} K \} dt. \\
\end{split}
\end{equation*}
The graph of the expected hedging error as a function with respect to variables strike price $K$ and Hurst parameter $H$ is plotted in below which points out that the strike price $K=1$ is a critical point. The expected hedging error tends to zero as strike price becomes bigger and bigger.

\begin{figure}[h!]
\begin{center}
\includegraphics[width=5.50cm]{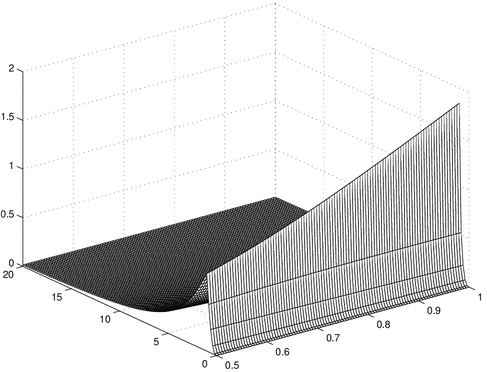}
\caption{Expected hedging error}
\end{center}
\end{figure}

\subsection{Asymptotic behavior with respect Hurst parameter H}

Assume $ B^{H}$ be a standard fractional Brownian motion with Hurst parameter $ H \in (0,1)$. It is straightforward to check that when the Hurst parameter $H$ tends to $H_0 \in (0,1)$, the finite-dimensional distributions of fractional Brownian motions $ \{ B^H \}$ tend to finite-dimensional distributions of $B^{H_0} $. It follows because of convergence of covariance functions and by
\begin{equation*}
 \E |B^{H}_t - B^{H}_s| ^ 2 = |t-s|^{2H}
\end{equation*}
and Billingsley criterion (see \cite{b}) we conclude that the family of laws of the fractional Brownian motions $\{ B^H \}$ converge in law in the space $C[0,T]$ of continuous function on the interval $[0,T]$ to that of $B^{H_0}$.\\
\vskip0.25cm
Moreover, the following theorem shows that the same holds for the family of local times $\{ l^{H}, H\in (0,1) \}$ of the fractional Brownian motions $\{ B^H \}$.

\begin{thm}\cite{j-v}\label{thm:local}
The family $\{ l^{H}, H\in (0,1) \}$ of local times of the fractional Brownian motions $\{ B^H, H\in(0,1) \}$ converges in law in the space $C([-D,D] \times [0,T])$, for any $D,T > 0$, to the local time $l^{H_0}$ of fractional Brownian motion $B^{H_0}$, when $H$ tends to $H_0$.
\end{thm}
\vskip0.10cm
Now consider the family of fractional Brownian motions $\{ B^H, H\in(\frac{1}{2},1) \} $. Denote Brownian motion $B$ and its local time $l(x,t)$ as the limit in law when $H$ tends to $\frac{1}{2} $ of corresponding fractional Brownian motions $\{ B^H \}$ and its local times. Let $\tilde{S}= S_0 \exp\{B\}$ be geometric Brownian motion and $ f(\tilde{S})$ be the payoff of European call option in Black--Scholes market model. Then we have that 

\begin{corl}
Assume the fixed proportional transaction costs coefficient $k_0 = \sqrt{\frac{\pi}{2}}$. Then
\begin{equation*}
 \lim_{H \downarrow \frac{1}{2}} \big( \lim_{n \to \infty} V_{\cdot}( {\theta}^{n} ) \big) = f(\tilde{S}_{\cdot}) - K l(\ln K,\cdot),
\end{equation*}
which limits take place in law in the space $C[0,T]$.
\end{corl}

\begin{corl}
Assume the fixed proportional transaction costs coefficient $k_0 = \frac{1}{2} \sqrt{\frac{\pi}{2}}$. Then
\begin{equation*}
 \lim_{n \to \infty} \big( \lim_{H \downarrow \frac{1}{2}} V_{\cdot}( {\theta}^{n} ) \big) = f(\tilde{S}_{\cdot}) - K l(\ln K,\cdot),
\end{equation*}
which limits again take place in law in the space $C[0,T]$.
\end{corl}

\subsection{Conclusion}

It is known that geometric fractional Brownian motion with Hurst parameter $H \in (\frac{1}{2},1)$ has zero quadratic variation process. Cheridito \cite{ch} uses this fact to show pricing model based on it, admits arbitrage. Moreover in our frictionless model in the case of European call option the formula (\ref{eq:chain}) implies that
\begin{equation*}
 (S_T-K)^+ = (S_0-K)^+ + \int _0 ^T 1 _{\{ S_t > K\} } d S_t.
\end{equation*}
This indicates that out-of-money options are worthless. Moreover the hedging strategy $u_t=1 _{\{ S_t > K\} } $ is an arbitrage opportunity. On the other hand, consider a frictionless pricing model with continuous price process  
\begin{equation*}
 X_t = S_0 \exp \{ B^{H}_t + \varepsilon W_t \} \quad \varepsilon >0,
\end{equation*}
where $W$ is a standard Brownian motion independent of $B^H$. This process has non-zero quadratic variation and fullfills all conditions of a recent result by Bender et.al. \cite{b-s-v}. Their result asserts that, this pricing model does not admit arbitrage opportunities with reasonable trading strategies. This indicates that the existence of non-zero quadratic variation is important for option pricing based on no arbitrage.\\

Let $f$ be a convex function with positive Radon measure $\mu$ as its second derivative. Consider continuous semimartingale $X_t = X_0 e ^{W_t}$ with local time $l_X$ and $X_0 \in \mathbb{R_{+}}$. By \emph{It\^o-Tanaka formula} (see \cite{r-y}, page 223) we have
\begin{equation*}
\begin{split}
 f(X_1) &= f(X_0) + \int _0 ^1 f^{'}_{-}(X_t) d X_t + \frac{1}{2} \int_{\mathbb{R}}l_{X} (a,[0,1]) \mu (da)\\
& = f(X_0) + \int _0 ^1 f^{'}_{-}(X_t) d X_t + \frac{1}{2} \int_{\mathbb{R}} a l_{W} (\ln a,[0,1]) \mu(da). 
\end{split}
\end{equation*}
Hence  our fractional Black-Scholes model with asymptotic proportional transaction costs has the same effect as the model which the stock price is modelled by semimartingale $X$. In this sense there is a connection between transaction costs and quadratic variation.

\textbf{Acknowledgements}.\\
Thanks are due to my supervisor Esko Valkeila for suggesting the problem of this paper and for helpful discussions. Also, I would like to thank Yuri Kabanov for useful comments and  Mario Wschebor for the reference \cite{a}. I am indebted to the Finnish Graduate School in  Stochastic and Statistic (FGSS) for financial support.


\begin{thebibliography}{9}
\bibitem{a} Aza\"is, J.M., \newblock \emph{ Conditions for convergence of number of crossings to the local time }. Probab. Math. Statist. \textbf{11}, no. 1, 19-36 (1990).

\bibitem{a-m-v} Azmoodeh, E., Mishura, Y., and Valkeila, E.,. \newblock \newblock\emph{ On hedging European options in geometric fractional Brownian motion market model}. Statistics \& Decisions, accepted (2010).


\bibitem{b-s-va} Bender, C., Sottinen, T., and Valkeila, E., \newblock \newblock\emph{ Fractional processes as models in stochastic finance }. To appear in Advanced Mathematical Methods for Finance (2010).

\bibitem{b-s-v} Bender, C., Sottinen, T., and Valkeila, E., \newblock \newblock\emph{ Pricing by hedging  beyond semimartingales}. Finance  Stoch, \textbf{12}, 441-468 (2008).

\bibitem{b} Billingsley, P., \newblock \emph{ Convergence of probability measures}. John Wiley \& Sons Inc., New York (1968).

\bibitem{b-h} Bj\"{o}rk, T., Hult, H., \newblock \emph{A note on wick products and the fractional black-scholes model},  Finance Stoch.  \textbf{9},  no. 2, 197-209 (2005). 

\bibitem{ch} Cheridito, P., \newblock\newblock \emph{ Arbitrage in fractional Brownian motion models}. Finance and Stochastics, \textbf{7}, 533-553 (2003).


\bibitem{c-n-w} Corcuera, J. M., Nualart, D., Woerner, H. C., \newblock \emph{Power variation of some integral fractional processes}.
Bernoulli \textbf{12}, no. 4, 713-735 (2006). 

\bibitem{g-h} Geman, D., Horowitz, J., \newblock \emph{Occupation densities}.  Ann. Probab.  \textbf{8}, no. 1, 1-67 (1980).

\bibitem{g} Guasoni, P., \newblock No arbitrage under transaction costs with fractional Brownian motion and beyond. \emph{ Math. Finance},\textbf{16}, 569-582 (2006).

\bibitem{g-r-s} Guasoni, P., Rásonyi, M., Schachermayer, W., \newblock \emph{Consistent price systems and face-lifting pricing under transaction costs}.  Ann. Appl. Probab. \textbf{18},  no. 2, 491-520 (2008).

\bibitem{g-r-s-1} Guasoni, P., Rásonyi, M., Schachermayer, W., \newblock \emph{The Fundamental Theorem of Asset Pricing for Continuous Processes under Small Transaction Costs}. Annals of Finance.  \textbf{6}, no. 2, 157-191 (2010). 

\bibitem{j-v}  Jolis, M., Viles, N., \newblock \emph{Continuity in Law with Respect to the Hurst Parameter of the Local Time of the Fractional Brownian Motion}. J. Theoret. Probab. \textbf{20}, no. 2, 133-152 (2007).

\bibitem{k-s} Kabanov, Y., Safarian, M., \newblock \newblock\emph{On Leland's strategy of option pricing with transaction costs}. Finance and Stochastic, \textbf{1}, 3, 239-250 (1997).

\bibitem{le} Leland, H., \newblock \newblock\emph{Option pricing and replication with transaction costs}. Journal of Finance, \textbf{XL}, 5, 1283-1301(1985).

\bibitem{lo} Lott, K., \newblock Ein Verfahren zur Replikation von Optionen unter Transaktionkosten in stetiger Zeit, Dissertation. \newblock \emph {Universit\"at der Bundeswehr M\"unchen. Institut f\"ur Mathematik und Datenverarbeitung} (1993).

\bibitem{m} Mishura, Y., \newblock \newblock\emph{Stochastic Calculus for Fractional Brownian Motion and Related Processes}, Lecture Notes in Mathematics, Vol. 1929, Springer, Berlin (2008).

\bibitem{pro} Protter, P., \newblock \emph{Stochastic integration and differential equations}. 2nd ed.,\newblock Springer, Berlin (2004).

\bibitem{r-y} Revuz, D., Yor. M., \newblock \newblock\emph{ Continuous martingales and Brownian motion}. Springer, Berlin (1999).

\bibitem{s-v} Sottinen, T. and Valkeila, E., \newblock \emph{ On arbitrage and replication in the fractional Black-Scholes pricing model}. Statistics \& Decisions, \textbf{21}, 93-107 (2003).

\bibitem{val} Valkeila, E., \newblock \emph{On the approximation of geometric fractional Brownian motion}. \newblock Optimality and Risk - Modern Trends in Mathematical Finance, The Kabanov Festschrift, 251-266 (2008).

\end{thebibliography}
\end{document}